\documentstyle[12pt,aasms4]{article}

\begin{document}

\title{The Ulysses Supplement to the GRANAT/WATCH Catalog
of Cosmic Gamma-Ray Bursts}

\author{K. Hurley}
\affil{University of California, Berkeley, Space Sciences Laboratory,
Berkeley, CA 94720-7450}

\author{N. Lund, S. Brandt}
\affil{Danish Space Research Institute, DK-2100 Copenhagen, Denmark}

\author{C. Barat}
\affil{CESR, F-31029 Toulouse Cedex, France}

\author{T. Cline}
\affil{NASA Goddard Space Flight Center, Greenbelt, MD 20771}

\author{R. Sunyaev, O. Terekhov, A. Kuznetsov, S. Sazonov}
\affil{IKI, 117810 Moscow, Russia}

\author{A. Castro-Tirado\altaffilmark{1}}
\affil{Laboratorio de Astrof\'{\i}sica Espacial y F\'{\i}sica 
Fundamenta (LAEFF-INTA), P.O. Box 50727, E-28080 Madrid, Spain}

\altaffiltext{1}{Instituto de Astrof\'{\i}sica de Andaluc\'{\i}a (IAA-CSIC),
P.O. Box 03004, E-18080 Granada, Spain}

\begin{abstract}

We present 3rd interplanetary network (IPN) localization data for 
56 gamma-ray bursts in the GRANAT/
WATCH catalog which occurred between 1990 November and 1994 September.  
These localizations are obtained by triangulation using various combinations
of spacecraft
in the IPN, which consisted of \it Ulysses \rm, BATSE,
\it Pioneer Venus Orbiter \rm (PVO), \it Mars Observer \rm (MO), WATCH, and PHEBUS.  The
intersections of the triangulation annuli with the WATCH error circles produce
error boxes with areas as small as 16 sq. arcmin., reducing the sizes of the
error circles by up to a factor of 800.

\end{abstract}


\section{Introduction}

The multi-wavelength counterparts to numerous gamma-ray bursts (GRBs) have
now been identified using the rapid, precise localizations available from
the BeppoSAX spacecraft (e.g. Costa et al. 1997; van Paradijs et al. 1997),
as well as from the \it Rossi X-Ray Timing Explorer \rm and the IPN.
However, there is still a need for less precise GRB localizations of older
bursts, for several reasons.  For example, the discovery of bright optical emission
coincident with one burst (Akerlof et al. 1999) indicates that searches through
archival optical data may reveal other examples of this interesting phenomenon.
Also, the possible association of one GRB with a nearby supernova (Galama et al.
1998), if valid, means that other such associations may exist in the historical
records.  Because the current rate of rapid, precise localizations remains low
($\sim$ 8 events/year), it is important to add as many bursts as possible to
the existing database.  The GRANAT/WATCH GRB catalog contains data on 95 bursts
observed between 1989 and 1994 (Sazonov et al. 1998); of the 95, 
47 bursts were localized to error circles with radii between 0.2 and 1.6 $\arcdeg$.  
The 3rd interplanetary network (IPN) began operations in 1990 with the launch
of the \it Ulysses \rm spacecraft.  By
combining WATCH data with IPN data, it
is possible to reduce the sizes of these error circles by as much as a
factor of 800, making
them more useful for archival studies.    This is the 7th in a series of catalogs 
of IPN localizations.  The supplements to the BATSE 3B and 4Br catalogs appeared
in Hurley et al. (1999a,b; 218 and 147 bursts, respectively).  Localizations involving the \it Mars Observer \rm (MO)
and \it Pioneer Venus Orbiter \rm (PVO) spacecraft have been presented in Laros et al. (1997,
1998; 9 and 37 bursts, respectively).  
Fifteen \it Ulysses \rm, PVO, SIGMA, WATCH, and PHEBUS burst localizations were
published in Hurley et al. (2000a).  \it Ulysses \rm/BeppoSAX bursts may be found
in Hurley et al. (2000b; 16 bursts).  Localization data for the bursts in all these
catalogs may also be found on the IPN website \footnote{ssl.berkeley.edu/ipn3/index.html}.

\section{Instrumentation}

The gamma-ray bursts in this paper were observed by at least two instruments.
One was the omnidirectional GRB detector aboard the \it Ulysses \rm spacecraft,
consisting of two 3 mm thick hemispherical CsI scintillators with a projected
area of $\sim$ 20 cm$^{2}$ in any direction.  The instrument observes bursts
in the 25 - 150 keV energy range in either a triggered mode, in which the
time resolution is as high as 8 ms, or, for the weaker bursts, in a real-time
mode, in which the time resolution is between 0.25 and 2 s.  A more complete
description of the experiment may be found in Hurley et al. (1992).

The second was the WATCH experiment aboard the GRANAT spacecraft.  WATCH
employs a unique rotating modulation collimator technique to determine the
positions of bursts to $\sim$ 1 $\arcdeg$ accuracy.  The detector is a scintillator
operating in the 8 - 60 keV range with a field of view of 74 $\arcdeg$ and a
maximum effective area of 47 cm$^2$.  Four independent modules were deployed
aboard the GRANAT spacecraft, and $\sim$80\% of the sky was monitored with
them.  See Sazonov et al. (1998) for a more detailed description.

To localize the GRBs in this supplement, use was sometimes made of the data
from other experiments, too.  These are noted in the following section.

\section{Technique}

The methodology employed here is similar or identical to that used for the
\it Ulysses \rm supplement to the BATSE 3B and 4B catalogs (Hurley et al. 1999a,b).  
Each WATCH burst was searched for in the \it Ulysses \rm data.  One or more annuli 
of possible arrival directions was derived by triangulation
for each burst identified using the data from \it Ulysses \rm and at least one 
other instrument.    The bursts in this catalog
thus fall into one of the following categories. \\

1. Event observed by \it Ulysses \rm and WATCH only.  In this case, the
triangulation annulus was obtained utilizing the data of these two instruments.\\

2. Event observed by \it Ulysses \rm, WATCH, and PHEBUS.  PHEBUS was also 
aboard the GRANAT spacecraft (Barat et al. 1988; Terekhov et al. 1991).  
It consisted
of six 12 cm. long by 7.8 cm. diameter BGO detectors oriented along the axes of a Cartesian
coordinate system, operating in the 100 keV - 100 MeV energy range, with 1/128 s to
1/32 s time resolution.  In this case,
the triangulation was done using \it Ulysses \rm and the instrument
which resulted in the most precise triangulation annulus.  The WATCH
data have the advantage of being taken in an energy range which corresponds
more closely to that of \it Ulysses \rm, but the time resolution of the WATCH
data was sometimes rather coarse ($\sim$ 10 s or more).  On the other hand,
the PHEBUS data, although taken in an energy range higher than that of
\it Ulysses \rm, have the advantages of good time resolution and in some
cases better statistics.  The more accurate of the two possible triangulation
annuli is quoted here. \\

3. Event observed by \it Ulysses \rm, WATCH, and BATSE.  BATSE consists
of eight detector modules
aboard the \it Compton Gamma-Ray Observatory \rm (GRO).  Each module has an area $\sim$
2025 cm$^2$.  The DISCSC data type was used, which gives 0.064 ms resolution data
for the 25-100 keV energy range.  BATSE is described in Meegan et al. (1996).   
For the purposes of triangulation, the GRANAT and GRO
spacecraft were close enough to one another ($<$ 250 light-ms) that the    
accuracy of the triangulation could not be improved by including the data
from both spacecraft.  (For comparison, the \it Ulysses \rm-Earth distance
was as great as several thousand light-seconds.)  In this case, the
\it Ulysses \rm - BATSE annulus was used, since the BATSE energy range corresponds
closely to that of \it Ulysses \rm, the time resolution is good, and the statistics
are always better.  These annuli have appeared in Hurley et al. (1999a), but
their intersections with the WATCH error circles are presented here for the
first time.  The BATSE error circles may be found in Meegan et al. (1996).
In those cases where WATCH did not localize the burst, the \it Ulysses \rm -
BATSE localization information consists of the intersection of the IPN
annulus with BATSE error circle.  Because the error circle is large, the
curvature of the annulus does not allow a simple description of error box,
and no localization information appears in table 2; it may be found in
Hurley et al. (1999a).

4. Event observed by \it Ulysses \rm, WATCH, and one or more of the
following experiments: BATSE, COMPTEL (Kippen et al. 1998),
SIGMA (Claret et al. 1994), PVO (Laros et al. 1997), or MO (Laros et al. 1998).
Here, triangulation using PVO or MO data, and/or the independent localization
capabilities of COMPTEL or SIGMA, 
have been utilized.  These special cases are
noted in table 2, and the previously published error box coordinates have been 
included in the table for convenience.  In most cases the error box is fully
contained within the WATCH error circle.
If no figure has been previously published showing the WATCH error circle and the IPN
triangulation result, one appears in this paper.

\section{The data}

In table 1 the WATCH bursts also detected by \it Ulysses \rm are listed.
Column 1 gives the date, column 2 gives the detection time at WATCH, and column
3 indicates the \it Ulysses \rm data mode (RI for rate increase, observed in
the low time resolution real-time mode, trigger for the high time resolution
triggered mode).  Column 4 indicates whether BATSE observed the burst.  Here
N/O means not observable (GRO had not been launched yet), and a number, if
present, is the BATSE trigger number.  Column 5 indicates whether the burst
was localized by WATCH, and column 6 indicates whether PHEBUS observed the
event.

Table 2 gives the localization information for the events in table 1.
Columns 1 and 2 give the date and the time.  For those bursts localized
by WATCH, columns 3 and 4 give
the right ascension and declination of the center of the WATCH error
circle (J2000), and column 5 gives the  WATCH 3 $\sigma$ error circle radius.
These data are taken directly from Sazonov et al. (1998).  Columns 6 and
7 give the right ascension and declination $\alpha, \delta$ of the center of the IPN
annulus (J2000); columns 8 and 9 give the radius R of the center line
of the annulus, and the 3 $\sigma$ half-width of the annulus $\delta$R.  That
is, the annulus is described by two small circles on the celestial sphere
both centered at $\alpha, \delta$, with radii R-$\delta$R and R+$\delta$R.
For those cases where there is a WATCH error circle and the annulus intersects
it, additional
data are given in columns 10 and 11.  (The possible exceptions are first,
cases where the IPN annulus is wider than the error circle diameter
and therefore does not intersect it, and second, cases where an actual error
box has been obtained and published elsewhere.)  Column 10 gives the right ascensions
and declinations (J2000) of the IPN error box and column 11 gives
the error box area.  Note that, strictly
speaking, it is not possible to define a true error box with straight line
segments between the four intersection points of a WATCH error circle with
an IPN annulus due to the curvatures of both
the annulus and the error circle.  However, for many purposes, this
may be negligible.

\section{Discussion and Conclusions}

There is good agreement between the IPN annuli and
the WATCH error circles in all cases.  We call attention to some of the more precise
error boxes:\\

1. 940703.  The error box area is 16 sq. arcmin., a reduction in area from
the 0.24$\arcdeg$ WATCH error circle of a factor of $\sim$41.\\

2. 921022.  The error box area is 22 sq. arcmin., a reduction in area from
the 0.72$\arcdeg$ WATCH error circle of a factor of $\sim$265.\\

3. 921013.  The error box area is 32 sq. arcmin., a reduction in area from
the 1.51$\arcdeg$ WATCH error circle of a factor of $\sim$800.\\

The localizations in table 2 are presented in figures 1-25.  (Figures for the
WATCH bursts involving SIGMA, which have already appeared in Hurley et al. (2000a),
have been omitted.)  As these figures show, the combination of WATCH and
the IPN results in very precise location information for these bursts.
Another version of the WATCH experiment was flown aboard the EURECA spacecraft.
Analysis of these events is currently underway.

\acknowledgments
KH is grateful to JPL for \it Ulysses \rm support under Contract 958056,
and to NASA for Compton Gamma-Ray Observatory support under
grant NAG 5-3811.

\newpage

\figcaption{IPN 3 $\sigma$ annulus and WATCH 3$\sigma$ error circle
for 901121.  The large width of the annulus, just visible in the
lower part of the figure, is due to the fact that
\it Ulysses \rm had just been launched, and was only 67 light-seconds
from Earth. \label{fig1}
}

\figcaption{IPN annulus (3 $\sigma$), COMPTEL (1 $\sigma$), WATCH 
(3 $\sigma$), and BATSE (1 $\sigma$) error circles for
910627.   \label{fig2}}

\figcaption{IPN annulus (3 $\sigma$), WATCH (3 $\sigma$), and BATSE (1 $\sigma$)
error circles for 910927.  \label{fig3}
}

\figcaption{PVO/\it Ulysses \rm / BATSE 3 $\sigma$  error box from Laros
et al. (1998), WATCH 3 $\sigma$
error circle, and BATSE 1 $\sigma$ error circle for 911202. \label{fig4}
}

\figcaption{IPN annulus (3 $\sigma$), WATCH (3 $\sigma$), and BATSE (1 $\sigma$)
error circles for 911209. \label{fig5}
}

\figcaption{PVO/\it Ulysses \rm / BATSE 3 $\sigma$  error box from Laros
et al. (1998), WATCH 3 $\sigma$
error circle, and BATSE 1 $\sigma$ error circle for 920311. \label{fig6}
}

\figcaption{PVO/\it Ulysses \rm / BATSE 3 $\sigma$  error box from Laros
et al. (1998), WATCH 3 $\sigma$
error circle, and BATSE 1 $\sigma$ error circle for 920404. \label{fig7}
}
\figcaption{IPN annulus (3 $\sigma$), WATCH (3 $\sigma$), and BATSE (1 $\sigma$)
error circles for the burst of 920718 at 14:40. \label{fig8}
}

\figcaption{IPN annulus (3 $\sigma$), WATCH (3 $\sigma$), and BATSE (1 $\sigma$)
error circles for the burst of 920718 at 21:32.\label{fig9}
}

\figcaption{PVO/\it Ulysses \rm / BATSE 3 $\sigma$  error box from Laros
et al. (1998), WATCH 3 $\sigma$
error circle, and BATSE 1 $\sigma$ error circle for 920720. \label{fig10}
}

\figcaption{IPN annulus (3 $\sigma$), WATCH (3 $\sigma$), and BATSE (1 $\sigma$)
error circles for 920814. \label{fig11}
}

\figcaption{IPN annulus (3 $\sigma$), WATCH (3 $\sigma$), and BATSE (1 $\sigma$)
error circles for 920902. \label{fig12}
}

\figcaption{IPN annulus (3 $\sigma$) and WATCH error circle (3 $\sigma$) for the
burst of 920903 at 01:37. \label{fig13}
}

\figcaption{IPN annulus (3 $\sigma$) and WATCH error circle (3 $\sigma$) for the
burst of 920903 at 23:29. \label{fig14}
}

\figcaption{IPN annulus (3 $\sigma$) and WATCH error circle (3 $\sigma$) for the
burst of 920925 at 20:30. \label{fig15}
}

\figcaption{IPN annulus (3 $\sigma$) and WATCH error circle (3 $\sigma$) for the
burst of 920925 at 22:46. \label{fig16}
}

\figcaption{IPN annulus (3 $\sigma$) and WATCH error circle (3 $\sigma$) for 
921013. \label{fig17}
}

\figcaption{IPN annulus (3 $\sigma$), BATSE (1 $\sigma$) and WATCH (3 $\sigma$)
error circles  for 
921022. \label{fig18}
}

\figcaption{IPN annulus (3 $\sigma$), BATSE (1 $\sigma$) and WATCH (3 $\sigma$) 
error circles  for 921029. \label{fig19}
}

\figcaption{IPN 3 $\sigma$ annuli (\it Ulysses \rm/BATSE/MO), and BATSE (1 $\sigma$),
COMPTEL (1 $\sigma$), and WATCH (3 $\sigma$) error circles for 930612.  The 
narrow inner IPN annulus and the wider outer IPN annulus
intersect outside the figure to form a long error box. \label{fig20}
}

\figcaption{IPN 3 $\sigma$ annulus and WATCH 3 $\sigma$ error circle for 930703.
\label{fig21}
}

\figcaption{IPN 3 $\sigma$ annuli (\it Ulysses \rm/BATSE/MO), BATSE (1 $\sigma$),
and WATCH (3 $\sigma$) error circles for 930706.  \label{fig22}
}

\figcaption{IPN annulus (3 $\sigma$), WATCH (3 $\sigma$), and BATSE (1 $\sigma$)
error circles for 940419.\label{fig23}
}

\figcaption{IPN annulus (3 $\sigma$), WATCH (3 $\sigma$), and BATSE (1 $\sigma$)
error circles for the burst of 940701.\label{fig24}
}

\figcaption{IPN annulus (3 $\sigma$), WATCH (3 $\sigma$), and BATSE (1 $\sigma$)
error circles for the burst of 940703.\label{fig25}
}

\clearpage

\begin{deluxetable}{cccccc}
\tablecaption{\it Ulysses - WATCH bursts.}
\tablehead{
\colhead{Date} & \colhead{Time} & \colhead{\it Ulysses \rm} 
& \colhead{BATSE} & \colhead{WATCH} & \colhead{PHEBUS} \\
}

\startdata

1990 Nov 12	&	14:57:41	&	RI	&	N/O	&	no	&	yes	\\
1990 Nov 21	&	18:07:11	&	RI	&	N/O	&	yes	&	yes	\\
1991 Jan 17	&	0:58:13	&	RI	&	N/O	&	no	&	yes	\\
1991 Jan 22	&	15:14:00	&	trigger	&	N/O	&	yes	&	yes	\\
1991 Feb 19	&	11:45:24	&	trigger	&	N/O	&	yes	&	N/O	\\
1991 Mar 10	&	13:02:03	&	trigger	&	N/O	&	yes	&	N/O	\\
1991 Apr 25	&	0:38:05	&	RI	&	109	&	no	&	no	\\
1991 May 17	&	5:02:38	&	RI	&		&	no	&	yes	\\
1991 Jun 27	&	4:29:23	&	trigger	&	451	&	yes	&	yes	\\
1991 Jul 17	&	4:33:06	&	trigger	&	543	&	no	&	yes	\\
1991 Jul 17	&	13:07:23	&	trigger	&		&	no	&	N/O	\\
1991 Jul 21	&	19:30:17	&	RI	&	563	&	no	&	no	\\
1991 Aug 14	&	19:14:38	&	RI	&	678	&	no	&	yes	\\
1991 Aug 15	&	12:34:25	&	RI	&		&	no	&	no	\\
1991 Sep 27	&	23:27:00	&	trigger	&	829	&	yes	&	N/O	\\
1991 Oct 16	&	11:01:34	&	RI	&	907	&	yes	&	yes	\\
1991 Oct 22	&	4:14:00	&	RI	&	914	&	no	&	no	\\
1991 Dec 2	&	20:28:51	&	trigger	&	1141	&	yes	&	N/O	\\
1991 Dec 9	&	18:36:11	&	trigger	&	1157	&	yes	&	N/O	\\
\tablebreak
1992 Mar 7	&	0:18:11	&	RI	&	1467	&	no	&	no	\\
1992 Mar 11	&	2:20:26	&	trigger	&	1473	&	yes	&	yes	\\
1992 Mar 25	&	17:17:37	&	trigger	&	1519	&	no	&	yes	\\
1992 Apr 4	&	13:11:45	&	trigger	&	1538	&	yes	&	N/O	\\
1992 Jul 11	&	16:09:17	&	trigger	&	1695	&	no	&	N/O	\\
1992 Jul 14	&	13:04:33	&	RI	&	1698	&	yes	&	yes	\\
1992 Jul 18	&	14:40:43	&	RI	&	1708	&	yes	&	no	\\
1992 Jul 18	&	21:32:44	&	trigger	&	1709	&	yes	&	yes	\\
1992 Jul 20	&	5:53:20	&	RI	&	1712	&	yes	&	no	\\
1992 Jul 23	&	1:00:49	&	trigger	&	1721	&	no	&	yes	\\
1992 Jul 23	&	20:03:09	&	trigger	&		&	yes	&	yes	\\
1992 Aug 14	&	6:10:35&	RI	&	1815	&	yes	&	no	\\
1992 Sep 2	&	0:29:02	&	trigger	&	1886	&	yes	&	yes	\\
1992 Sep 3	&	1:35:46	&	RI    	&		&	yes	&	yes	\\
1992 Sep 3	&	23:29:01	&	trigger	&		&	yes	&	yes	\\
1992 Sep 25	&	20:30:42	&	RI	&		&	yes	&	yes	\\
1992 Sep 25	&	21:45:20	&	RI	&	1956	&	no	&	no	\\
1992 Sep 25	&	22:46:24	&	RI	&		&	yes	&	no	\\
1992 Oct 13	&	23:00:42	&	trigger	&		&	yes	&	yes	\\
\tablebreak
1992 Oct 22	&	15:21:00	&	trigger	&	1997	&	yes	&	N/O	\\
1992 Oct 25	&	13:55:40	&	trigger	&		&	no	&	yes	\\
1992 Oct 29	&	12:38:05	&	RI	&	2018	&	yes	&	no	\\
1993 Jan 6	&	15:37:40	&	trigger	&	2121	&	no	&	yes	\\
1993 Jan 16	&	2:47:06	&	RI	&	2136	&	no	&	no	\\
1993 Jun 9	&	10:07:30	&	RI	&	2383	&	no	&	no	\\
1993 Jun 12	&	0:44:20	&	RI	&	2387	&	yes	&	yes	\\
1993 Jul 3	&	11:26:30	&	trigger	&		&	yes	&	N/O	\\
1993 Jul 5	&	12:39:18	&	RI	&	2429	&	no	&	no	\\
1993 Jul 6	&	5:13:31	&	trigger	&	2431	&	yes	&	yes	\\
1993 Jul 14	&	16:13:04	&	RI	&	2446	&	no	&	yes	\\
1993 Sep 10	&	12:12:30	&	RI	&	2522	&	no	&	no	\\
1993 Sep 27	&	4:18:15	&	RI	&	2542	&	no	&	no	\\
1994 Mar 29	&	18:15:44	&	RI	&	2897	&	no	&	no	\\
1994 Apr 19	&	19:11:07	&	RI	&	2940	&	yes	&	yes	\\
1994 Jun 19	&	21:32:32	&	RI	&	3035	&	no	&	yes	\\
1994 Jul 1	&	21:44:29	&	RI	&	3055	&	yes	&	no	\\
1994 Jul 3	&	4:40:55	&	trigger	&	3057	&	yes	&	yes	\\
1994 Sep 10	&	23:57:56	&	trigger	&		&	no	&	yes	\\

\enddata

\end{deluxetable}

\clearpage
\begin{deluxetable}{ccccccccccc}
\scriptsize
\tablecaption{\it Localizations.}
\tablehead{
\colhead{Date} & \colhead{Time} & \multicolumn{3}{c}{WATCH} & \multicolumn{4}{c}{IPN}
& {Error box corners} & \colhead{Area} \\
\cline{3-5} \cline{6-9} \\
\colhead{}     & \colhead{}    & \colhead{$\alpha$} & \colhead{$\delta$} &
\colhead{R} & \colhead{$\alpha$} & \colhead{$\delta$} &
\colhead{R} & \colhead{$\delta$R} & \colhead{$\alpha,\delta$} &
\colhead{Sq. arcmin.} \\
}

\startdata

1990 Nov 12&14:57:41&	&	&	&115.525&28.474&48.575&0.678&	&	\\
1990 Nov 21&18:07:11&30.39&72.40&0.59&113.042&28.773&61.102&2.369&	&	\\
1991 Jan 17&0:58:13&	&	&	&95.972&27.621&64.757&0.629&	&	\\
1991 Jan 22\tablenotemark{a}&15:14:00&297.48&-71.23&0.69&	& & & &296.918,-70.681 &	\\
                            &        &      &      &    &   & & & &296.595,-70.612 & \\
                            &        &      &      &    &   & & & &296.674,-70.660 & \\
                            &        &      &      &    &   & & & &296.838,-70.633 & \\
                            &        &      &      &    &   & & & &297.000,-70.667 & \\
                            &        &      &      &    &   & & & &296.512,-70.626 & 18    \\
1991 Feb 19\tablenotemark{a}&11:45:24&212.94&58.54&0.95&	& & & &213.731,58.671 &	\\
                            &        &      &      &    &   & & & &213.657,58.705 &     \\
                            &        &      &      &    &   & & & &213.723,58.710 &     \\
                            &        &      &      &    &   & & & &213.665,58.666 &     \\
                            &        &      &      &    &   & & & &213.701,58.649 &     \\
                            &        &      &      &    &   & & & &213.687,58.727 & 7.3   \\
1991 Mar 10\tablenotemark{a}&13:02:03&184.10&6.38&0.55&     & & & &184.358,7.266 &	\\
                            &        &      &    &    &     & & & &184.249,7.125 &    \\
                            &        &      &    &    &     & & & &184.198,6.921 &    \\
                            &        &      &    &    &     & & & &184.405,7.462 &    \\
                            &        &      &    &    &     & & & &184.424,7.480 &    \\
                            &        &      &    &    &     & & & &184.178,6.901 & 63  \\
1991 Apr 25\tablenotemark{b}&0:38:05&	& & &	& & &	& & 	\\
1991 May 17\tablenotemark{a}&5:02:38&       &    &    &	& & & &150.475,-42.876 & 	\\
                            &        &      &    &    &     & & & &150.730,-42.693 &    \\
                            &        &      &    &    &     & & & &149.659,-43.107 &    \\
                            &        &      &    &    &     & & & &151.546,-42.447 &    \\
                            &        &      &    &    &     & & & &151.545,-42.447 &     \\
                            &        &      &    &    &     & & & &149.659,-43.107 &236     \\
1991 Jun 27\tablenotemark{c,d}&4:29:23&199.60&-2.60&1.09&134.826&18.423&66.820&0.012&198.986,-3.502  & \\
                            &       &      &     &    &       &      &      &     &199.758,-1.522  & \\
                            &       &      &     &    &       &      &      &     &198.966,-3.488  & \\
                            &       &      &     &    &       &      &      &     &199.734,-1.518  & 180.	\\
\tablebreak
1991 Jul 17\tablenotemark{c}&4:33:06&	& & &	& &	& &247.4875,-59.2052& 	\\
                            &       &     & & & & &   & &247.2591,-58.1126& \\
                            &       &     & & & & &   & &247.3284,-58.3370& \\
                            &       &     & & & & &   & &247.4138,-58.9874& \\
                            &       &     & & & & &   & &247.4810,-59.2876& \\
                            &       &     & & & & &   & &247.2667,-58.0269& 66. \\
1991 Jul 17&13:07:23& & &	&320.805&-16.587&89.903&0.252&	&	\\
1991 Jul 21\tablenotemark{b}&19:30:17&	& & &	& &	& & &	\\
1991 Aug 14\tablenotemark{c}&19:14:38&	& & &	& & &	&344.4255,29.0772 & 	\\
                            &       &     & & & & &   & &343.2790,29.4813& \\
                            &       &     & & & & &   & &343.5043,29.3878& \\
                            &       &     & & & & &   & &344.1460,29.1948& \\
                            &       &     & & & & &   & &345.2056,28.8168& \\
                            &       &     & & & & &   & &342.7718,29.6208& 139. \\
1991 Sep 27\tablenotemark{b}&23:27:00&49.70&-42.72&0.94&338.937&-10.074&68.870&0.036&49.326,-43.620 & \\
         &        &     &      &    &       &       &      &     &49.579,-41.784 & \\
         &        &     &      &    &       &       &      &     &49.230,-43.595 & \\
         &        &     &      &    &       &       &      &     &49.482,-41.794 &475. \\
1991 Oct 16\tablenotemark{a}&11:01:34&297.37&-4.71&0.92&   &   & & & 297.996,-5.386 & 	\\
                            &        &      &     &    &   &   & & & 298.151,-4.220 &      \\
                            &        &      &     &    &   &   & & & 298.148,-5.205 &      \\
                            &        &      &     &    &   &   & & & 298.251,-4.434 & 540  \\
1991 Oct 22\tablenotemark{b}&4:14:00&	& & &	& & &	& &	\\
1991 Dec 2\tablenotemark{c}&20:28:51&171.97&-22.59&0.94&	& & &	&171.6210,-23.3936 &  \\
                           &        &      &      &    &    & & & &173.9107,-22.9482 &  \\
                           &        &      &      &    &    & & & &173.7281,-23.0788 & \\
                           &        &      &      &    &    & & & &171.8002,-23.2773 & \\
                           &        &      &      &    &    & & & &171.6010,-23.3436 & \\
                           &        &      &      &    &    & & & &173.9318,-22.9972 & 707. \\
1991 Dec 9\tablenotemark{b}&18:36:11&261.92&-44.19&0.78&348.333&-6.380&82.576&0.003&262.465,-44.866 & \\
        &        &      &      &    &       &      &      &     &262.520,-43.541 & \\
        &        &      &      &    &       &      &      &     &262.474,-44.863 & \\
        &        &      &      &    &       &      &      &     &262.529,-43.545 &29 \\
\tablebreak
1992 Mar 7\tablenotemark{b}&0:18:11&	& & &	& & &	& &		\\
1992 Mar 11\tablenotemark{c}&2:20:26&132.25&-36.39&0.33&	& & &	&131.9906,-36.3137 & 	\\
                           &        &      &      &    &    & & & &132.3481,-36.4731 &   \\
                           &        &      &      &    &    & & & &132.1533,-36.3975 & \\
                           &        &      &      &    &    & & & &132.1847,-36.3899 & \\
                           &        &      &      &    &    & & & &132.0761,-36.3403 & \\
                           &        &      &      &    &    & & & &132.2621,-36.4469 & 15. \\
1992 Mar 25\tablenotemark{c}&17:17:37&	& & &	& & &	&350.5032,13.0873 & 	\\
                            &       &     & & & & &   & &350.6150,13.0463& \\
                            &       &     & & & & &   & &350.5214,13.0739& \\
                            &       &     & & & & &   & &350.5968,13.0597& \\
                            &       &     & & & & &   & &350.5757,13.0725& \\
                            &       &     & & & & &   & &350.5425,13.0612&4.8 \\
1992 Apr 4\tablenotemark{c}&13:11:45&323.07&22.53&0.66&	& & &	&323.2934,22.4946 & 	\\
                           &        &      &      &    &    & & & &323.4784,22.5744 & \\
                           &        &      &      &    &    & & & &323.4180,22.5292 & \\
                           &        &      &      &    &    & & & &323.3537,22.5400 & \\
                           &        &      &      &    &    & & & &323.3093,22.5176 & \\
                           &        &      &      &    &    & & & &323.4625,22.5515 & 14.1 \\
1992 Jul 11\tablenotemark{c}&16:09:17&	& & &	& & &	&281.5245,72.8744 & 	\\
                            &       &     & & & & &   & &281.5087,72.8867& \\
                            &       &     & & & & &   & &281.5545,72.8851& \\
                            &       &     & & & & &   & &281.4786,72.8760& \\
                            &       &     & & & & &   & &281.4663,72.8691& \\
                            &       &     & & & & &   & &281.5669,72.8919&0.92 \\
1992 Jul 14\tablenotemark{a}&13:04:33&221.43&-30.75&0.52&	& & & & 220.826,-30.721 & 	\\
                            &        &      &      &    &   & & & & 220.897,-30.506 &  \\
                            &        &      &      &    &   & & & & 220.848,-30.607 & 36 \\
1992 Jul 18\tablenotemark{b}&14:40:43&21.37&-3.36&0.78&332.718&-4.782&49.196&0.032&22.079,-3.688 & \\
         &        &     &     &    &       &      &      &     &22.034,-2.949 & \\
         &        &     &     &    &       &      &      &     &22.019,-3.794 & \\
         &        &     &     &    &       &      &      &     &21.963,-2.852 &175 \\
\tablebreak
1992 Jul 18\tablenotemark{b}&21:32:44&296.17&-55.95&0.65&332.751&-4.760&58.948&0.004&296.924,-56.446 & \\
         &        &      &      &    &       &      &      &     &295.262,-55.549 & \\
         &        &      &      &    &       &      &      &     &296.936,-56.441 & \\
         &        &      &      &    &       &      &      &     &295.271,-55.542 &39 \\
1992 Jul 20\tablenotemark{c}&5:53:20&145.67&-11.20&1.09&	& & &	&145.5853,-11.0979 & 	\\
                           &        &      &      &    &    & & & &145.3282,-10.4052 & \\
                           &        &      &      &    &    & & & &145.0168,-10.8227 & \\
                           &        &      &      &    &    & & & &145.8982,-10.6772 & \\
                           &        &      &      &    &    & & & &146.0366,-10.8810 & \\
                           &        &      &      &    &    & & & &144.8843,-10.6118 & 1580. \\
1992 Jul 23\tablenotemark{b}&1:00:49&	& & &	& & &	& & 	\\
1992 Jul 23\tablenotemark{a}&20:03:09&287.08&27.33&0.28&	& & &	&287.128,27.216 & 	\\
                            &        &      &     &    &    & & & &287.155,27.248  &     \\
                            &        &      &     &    &    & & & &287.126,27.210  &     \\
                            &        &      &     &    &    & & & &287.157,27.255  &     \\
                            &        &      &     &    &    & & & &287.148,27.249  &     \\
                            &        &      &     &    &    & & & &287.135,27.215  & 0.9 \\
1992 Aug 14\tablenotemark{b}&6:10:35&259.83&-45.17&1.24&336.053&-2.478&78.143&0.006&260.657,-46.267 & \\
         &       &      &      &    &       &      &      &     &260.166,-43.953 & \\
         &       &      &      &    &       &      &      &     &260.672,-46.262 & \\
         &       &      &      &    &       &      &      &     &260.182,-43.956 & 95 \\
1992 Sep 2\tablenotemark{b}&0:29:02&279.08&-22.81&0.46&338.594&-0.651&61.607&0.015&279.400,-23.163 & \\
        &       &      &      &    &       &      &      &     &279.212,-22.366 & \\
        &       &      &      &    &       &      &      &     &279.427,-23.141 & \\
        &       &      &      &    &       &      &      &     &279.246,-22.376 &86	\\
1992 Sep 3&1:35:46&295.87&35.46&0.70&338.740&-0.540&54.168&0.215&295.021,35.571 & \\
        &       &      &     &    &       &      &      &     &295.380,36.036 & \\
        &       &      &     &    &       &      &      &     &295.215,35.009 & \\
        &       &      &     &    &       &      &      &     &296.093,36.136 & 855	\\
\tablebreak
1992 Sep 3&23:29:01&301.54&22.59&0.30&338.863&-0.453&43.062&0.007&301.344,22.351 & \\
        &        &      &     &    &       &      &      &     &301.638,22.876 & \\
        &        &      &     &    &       &      &      &     &301.356,22.343 & \\
        &        &      &     &    &       &      &      &     &301.652,22.872 & 29	\\
1992 Sep 25&20:30:42&201.11&42.20&0.76&161.767&-1.864&56.582&0.007&201.776,41.624 & \\
         &        &      &     &    &       &      &      &     &200.440,42.778 & \\
         &        &      &     &    &       &      &      &     &201.762,41.615 & \\
         &        &      &     &    &       &      &      &     &200.426,42.769 &74	\\
1992 Sep 25\tablenotemark{b}&21:45:20& & &	& & &	& & 	\\
1992 Sep 25&22:46:24&330.80&25.48&0.39&341.779&1.878&25.717&0.086&330.448,25.255 & \\
         &        &      &     &    &       &     &      &     &331.214,25.591 & \\
         &        &      &     &    &       &     &      &     &330.603,25.133 & \\
         &        &      &     &    &       &     &      &     &331.224,25.404 &475	\\
1992 Oct 13&23:00:42&117.71&33.41&0.32&163.948&-3.879&57.350&0.007&117.556,33.117 & \\
         &        &     &    &    &       &      &      &     &117.950,33.660 & \\
         &        &     &    &    &       &      &      &     &117.572,33.112 & \\
         &        &     &    &    &       &      &      &     &117.963,33.651 &32 \\
1992 Oct 22\tablenotemark{b}&15:21:00&254.43&-9.64&0.72&164.862&-4.861&88.377&0.002&254.093,-10.279 & \\
         &        &      &     &    &       &      &      &     &253.993,-9.063 & \\
         &        &      &     &    &       &      &      &     &254.088,-10.276 & \\
         &        &      &     &    &       &      &      &     &253.988,-9.067 &22	\\
1992 Oct 25&13:55:40& & &	&345.153&5.208&27.696&0.013&	&	\\
1992 Oct 29\tablenotemark{b}&12:38:05&35.82&-0.42&1.25&345.514&5.651&50.169&0.045&35.277,-1.546 & \\
         &        &     &     &    &       &     &      &     &35.594,0.809 & \\
         &        &     &     &    &       &     &      &     &35.192,-1.501 & \\
         &        &     &     &    &       &     &      &     &35.501,0.789 &770 \\
1993 Jan 6\tablenotemark{b}&15:37:40&	& & &	& & &	& & \\
1993 Jan 16\tablenotemark{b}&2:47:06&	& & &	& & &	& &	\\
1993 Jun 9\tablenotemark{b}&10:07:30&	& & &	& & &	& & 	\\
\tablebreak
1993 Jun 12\tablenotemark{d,e}&0:44:20&109.24&-71.20&0.70&143.975&-11.183&63.927&0.053&110.227,-71.826 & \\
                 &       &      &      &    &       &       &      &     &107.144,-71.028 & \\
                 &       &      &      &    &       &       &      &     &110.533,-71.767 & \\
                 &       &      &      &    &       &       &      &     &107.261,-70.922 &480\\
1993 Jul 3&11:26:30&311.06&8.04&0.77&326.006&11.775&15.185&0.044&311.244,7.292&	\\
          &        &      &    &    &       &      &      &     &310.830,8.776&     \\
          &        &      &    &    &       &      &      &     &311.328,7.317&     \\
          &        &      &    &    &       &      &      &     &310.915,8.797& 480 \\
1993 Jul 5\tablenotemark{b}&12:39:18&	& & &	& & &	& &	\\
1993 Jul 6\tablenotemark{e}&5:13:31&281.42&-20.18&0.47&	& & &	&281.2524,-20.1137 & 	\\
                           &        &      &      &    &    & & & &281.2359,-20.0678 &  \\
                           &        &      &      &    &    & & & &281.2214,-20.0669 & \\
                           &        &      &      &    &    & & & &281.2669,-20.1146 & \\
                           &        &      &      &    &    & & & &281.2674,-20.1259 & \\
                           &        &      &      &    &    & & & &281.2209,-20.0557 & 3 \\
1993 Jul 14\tablenotemark{b}&16:13:04&	& & &	& & &	& & 	\\
1993 Sep 10\tablenotemark{b}&12:12:30&	& & &	& & &	& & 	\\
1993 Sep 27\tablenotemark{b}&4:18:15&	& & &	& & &	& & 	\\
1994 Mar 29\tablenotemark{b}&18:15:44&	& & &	& & &	& & 	\\
1994 Apr 19\tablenotemark{b}&19:11:07&358.82&-48.19&1.02&121.768&-46.270&73.314&0.062&357.496,-48.709 & \\
         &        &      &      &    &       &       &      &     &359.915,-47.483 & \\
         &        &      &      &    &       &       &      &     &357.601,-48.813 & \\
         &        &      &      &    &       &       &      &     &0.036,-47.577 & 906 \\
1994 Jun 19\tablenotemark{b}&21:32:32&	& & &	& & &	& & 	\\
\tablebreak
1994 Jul 1\tablenotemark{b}&21:44:29&145.67&-6.15&1.56&126.479&-39.199&37.139&0.157&147.078,-6.841 & \\
        &        &      &     &    &       &       &      &     &144.231,-5.530 & \\
        &        &      &     &    &       &       &      &     &146.909,-7.108 & \\
        &        &      &     &    &       &       &      &     &144.135,-5.830 & 3500 \\
1994 Jul 3\tablenotemark{b}&4:40:55&133.20&28.11&0.24&126.771&-39.239&67.508&0.005&133.434,27.988 & \\
        &       &      &     &    &       &       &      &     &132.944,28.029 & \\
        &       &      &     &    &       &       &      &     &133.427,27.978 & \\
        &       &      &     &    &       &       &      &     &132.949,28.018 &16	\\
1994 Sep 10&23:57:56&	&	&	&150.200&-51.597&70.732&0.046	& & 	\\

\enddata
\tablenotetext{a} {Localized using PVO, SIGMA, and PHEBUS (Hurley et al. 2000)}
\tablenotetext{b} {\it Ulysses \rm-BATSE error box in Hurley et al. (1999a)}
\tablenotetext{c} {Also observed by and/or localized using PVO (Laros et al. 1998)}
\tablenotetext{d} {Also localized by COMPTEL (Kippen et al. 1998)}
\tablenotetext{e} {Also observed by and/or localized using MO (Laros et al. 1997)}

\end{deluxetable}

\end{document}